\documentclass[12pt]{article}
\begin{document}

\begin{center}
{\Large\bf{}The modification of the Einstein and Landau-Lifshitz
pseudotensrs}
\end{center}

\begin{center}
Lau Loi So\\
Department of Physics, Tamkang University, Tamsui 251, Taiwan.
\end{center}

\begin{abstract}
Deser $et$ $al$. proposed a combination of the Einstein and
Landau-Lifshitz pseudotensors such that the second derivatives in
vacuum are proportional to the Bel-Robinson tensor. Stimulated by
their work, the present paper discuss the gravitational
energy-momentum expression which has the same desirable
Bel-Robinson tensor property.  We find modifications of the
Einstein and Landau-Lifshitz pseudotensors that both give the same
coefficient of the Bel-Robinson tensor in vacuum in holonomic
frames.
\end{abstract}

\section{Introduction}
The classical pseudotensor is not a tensorial object, it is frame
dependent.  A desirable requirement is that to the lowest
non-vanishing order in vacuum the energy density for a gravitating
system should be proportional to the Bel-Robinson tensor
$B_{\alpha\beta\mu\nu}$ \cite{Szabados}. This will assume a
covariant and positive energy on all reference frames, i.e. the
associated energy-momentum vector is future pointing and
non-spacelike.  We find one expression associated with a good, but
more complicated boundary condition which gives the Bel-Robinson
tensor only \cite{So2}. However, it is not easy to obtain the
Bel-Robinson tensor alone and in general it is accompanied by
certain other tensors,
$S_{\alpha\beta\mu\nu},~Y_{\alpha\beta\mu\nu}$ and
$T_{\alpha\beta\mu\nu}$ \cite{Deser}.\\

According to the standard textbook \cite{MTW}, the Einstein
pseudotensor does not give a positive gravitational energy in
vacuum to second order . Recently Deser $et$ $al$. \cite{Deser}
found the analogous expansion for the Landau-Lifshitz pseudotensor
by using the similar method, however this too does not have the
expected result. Yet these two classical pseudotensors are good,
giving physical sensible results, inside matter (mass density) and
at spatial infinity (ADM mass). There comes a natural question can
they be modified, so that they have the extra nice result of the
positive gravitational energy in vacuum? The answer is yes.\\

Moreover, the present paper gives the coefficient of the positive
gravitational energy for some suitable pseudotensors expressions
in holonomic frames, namely the combination of the Einstein and
Landau-Lifshitz pseudotensors \cite{Deser}, one of the Chen-Nester
four quasilocal expressions \cite{So2} and Papapetrou pseudotensor
\cite{So1}.

\section{Ingredients}

\subsection{Riemann normal coordinate}
In general the metric tensor $g_{\alpha\beta}$, using the Taylor
series expansion, can be expanded as
\begin{equation}
g_{\alpha\beta}(x)
=g_{\alpha\beta}(0)+(\partial_{\mu}g_{\alpha\beta})(0)x^{\mu}
+\frac{1}{2}(\partial^{2}_{\mu\nu}g_{\alpha\beta})(0)x^{\mu}x^{\nu}+\ldots
\end{equation}
At the origin in Riemann normal coordinates
\begin{eqnarray}
g_{\alpha\beta}(0)&=&\eta_{\alpha\beta},
\quad\quad\quad\quad\quad\quad\quad\quad~~
\partial_{\mu}g_{\alpha\beta}(0)=0, \\
-3\partial^{2}_{\mu\nu}g_{\alpha\beta}(0)
&=&R_{\alpha\mu\beta\nu}+R_{\alpha\nu\beta\mu},\quad\quad
-3\partial_{\nu}\Gamma^{\mu}{}_{\alpha\beta}(0)
=R^{\mu}{}_{\alpha\beta\nu}+R^{\mu}{}_{\beta\alpha\nu}.
\end{eqnarray}
Hence
\begin{equation}\label{7a Nov 2005}
g_{\alpha\beta}(x)=\eta_{\alpha\beta}
-\frac{1}{3}R_{\alpha\xi\beta\kappa}x^{\xi}x^{\kappa}+O(x^{3}).
\end{equation}

\subsection{The tensors B, S, Y, T and their significance}
In order to extract the physical meaning of the tensors
$B_{\alpha\beta\mu\nu}$, $T_{\alpha\beta\mu\nu}$,
$S_{\alpha\beta\mu\nu}$ and $Y_{\alpha\beta\mu\nu}$, one can use
the analog of the \symbol{92}electric" $E_{ij}$ and
\symbol{92}magnetic" $H_{ij}$ parts in vacuum of the Weyl tensor,
\begin{eqnarray}
E_{ij}=C_{0i0j}, \quad H_{ij}=\ast{C_{0i0j}},
\end{eqnarray}
where $C_{\alpha\beta\mu\nu}$ is the Weyl conformal tensor and
$\ast{C_{\alpha\beta\mu\nu}}$ is its dual,
\begin{equation}
\ast{C_{\alpha\beta\mu\nu}}
=\frac{1}{2}\sqrt{-g}\epsilon_{\alpha\beta\lambda\sigma}
C^{\lambda\sigma}{}_{\mu\nu}.
\end{equation}
In a simple form using the Riemann tensor in vacuum
\begin{equation}
E_{ab}=R_{0a0b},\quad
H_{ab}=\frac{1}{2}R_{0amn}\epsilon_{b}{}^{mn}.
\end{equation}
Certain commonly occurring quadratic combinations of the Riemann
tensor components in terms of the electric $E_{ab}$ and magnetic
$H_{ab}$ parts in vacuum are
\begin{equation}
R_{0a0b}R_{0}{}^{a}{}_{0}{}^{b}=E_{ab}E^{ab},\quad
R_{0abc}R_{0}{}^{abc}=2H_{ab}H^{ab},\quad
R_{abcd}R^{abcd}=4E_{ab}E^{ab}.
\end{equation}
In particular, the Riemann squared tensor can be written in terms
of the electric and magnetic parts as
\begin{eqnarray}
R_{\alpha\beta\mu\nu}R^{\alpha\beta\mu\nu}
&=&4R_{0a0b}R^{0a0b}+4R_{0abc}R^{0abc}+R_{abcd}R^{abcd} \nonumber \\
&=&8(E_{ab}E^{ab}-H_{ab}H^{ab}),
\end{eqnarray}
where Greek letter means $0,1,2,3$ and Latin stands for $1,2,3$.
The Bel-Robinson tensor $B_{\alpha\beta\mu\nu}$ and tensor
$S_{\alpha\beta\mu\nu}$ are defined as follows
\begin{eqnarray}
B_{\alpha\beta\mu\nu}&:=&R_{\alpha\lambda\mu\sigma}R_{\beta}{}^{\lambda}{}_{\nu}{}^{\sigma}
+R_{\alpha\lambda\nu\sigma}R_{\beta}{}^{\lambda}{}_{\mu}{}^{\sigma}
-\frac{1}{8}g_{\alpha\beta}g_{\mu\nu}R_{\rho\tau\lambda\sigma}R^{\rho\tau\lambda\sigma},\\
S_{\alpha\beta\mu\nu}&:=&R_{\alpha\mu\lambda\sigma}R_{\beta\nu}{}{}^{\lambda\sigma}
+R_{\alpha\nu\lambda\sigma}R_{\beta\mu}{}{}^{\lambda\sigma}
+\frac{1}{4}g_{\alpha\beta}g_{\mu\nu}R_{\rho\tau\lambda\sigma}R^{\rho\tau\lambda\sigma}.
\end{eqnarray}
For our future analysis, define the tensors
$Y_{\alpha\beta\mu\nu}$ and $T_{\alpha\beta\mu\nu}$ \cite{Deser}
\begin{eqnarray}
Y_{\alpha\beta\mu\nu}
&:=&R_{\alpha\lambda\beta\sigma}R_{\mu}{}^{\lambda}{}_{\nu}{}^{\sigma}
+R_{\alpha\lambda\beta\sigma}R_{\nu}{}^{\lambda}{}_{\mu}{}^{\sigma},\\
T_{\alpha\beta\mu\nu}&:=&-\frac{1}{24}g_{\alpha\beta}g_{\mu\nu}
R_{\lambda\sigma\rho\tau}R^{\lambda\sigma\rho\tau}.
\end{eqnarray}
The physical observable of the energy-momentum for the tensors
$B_{\alpha\beta\mu\nu}$, $T_{\alpha\beta\mu\nu}$,
$S_{\alpha\beta\mu\nu}$ and $Y_{\alpha\beta\mu\nu}$ are the
spatial traces
\begin{eqnarray}
B_{\mu0l}{}^{l}&=&(E_{ab}E^{ab}+H_{ab}H^{ab},2\epsilon_{c}{}^{ab}E_{ad}H^{d}{}_{b}),\\
T_{\mu0l}{}^{l}&=&(E_{ab}E^{ab}-H_{ab}H^{ab},0),\\
S_{\mu0l}{}^{l}&=&-10T_{\mu0l}{}^{l},\\
Y_{\mu0l}{}^{l}&=&2(E_{ab}E^{ab},\epsilon_{c}{}^{ab}E_{ad}H^{d}{}_{b}).
\end{eqnarray}
Note that $B_{\mu{}0l}{}^{l}$ has a form similar to the Maxwell
energy-momentum density and in particular $B_{00l}{}^{l}\geq{}0$.
There is one relation which needs our attention throughout the
present text
\begin{equation}
B_{\mu{}0l}{}^{l}=S_{\mu{}0l}{}^{l}+Y_{\mu{}0l}{}^{l}+9T_{\mu{}0l}{}^{l}.
\end{equation}
This means that it is not necessary to obtain the Bel-Robinson
tensor $B_{\mu{}0l}{}{}^{l}$ for the positive energy requirement,
a combination of the tensors $Y_{\mu{}0l}{}^{l}$ with
$T_{\mu{}0l}{}^{l}$ or equivalently $Y_{\mu{}0l}{}^{l} $ with
$S_{\mu{}0l}{}^{l}$ can suffice.

\section{The interior, ADM and gravitational energy}
Note three physical quantities of interest if one considers a
massive object in general relativity.  They are the interior mass
density, the ADM mass \cite{ADM} at the spatial infinity and the
gravitational field energy-momentum in vacuum. In order to study
the gravitational energy, Einstein proposed the classical
pseudotensor $t_{\alpha}{}^{\mu}$ which follows from the
superpotential $U_{\alpha}{}^{[\mu\nu]}$. Unfortunately the
superpotential is not uniquely defined, for example
\begin{equation}
t_{\alpha}{}^{\mu}=\partial_{\nu}U_{\alpha}{}^{[\mu\nu]},
\end{equation}
but one can introduce a new pseudotensor such as
\begin{equation}
\widetilde{t}_{\alpha}{}^{\mu}=t_{\alpha}{}^{\mu}
+\partial_{\nu}\widetilde{U}_{\alpha}{}^{[\mu\nu]},
\end{equation}
which is likewise conserved may seem that there is no special way
to study the gravitational energy. However, the interior and ADM
mass provided some guidelines or restriction, so that one can have
some sort of physical energy-momentum components.  Recalling the
Freud superpotential, one may consider the generalization
\begin{equation}
U_{\alpha}{}^{[\mu\nu]}=\sqrt{-g}\left\{
k_{1}(\delta^{\mu}_{\alpha}\Gamma^{\lambda}{}_{\lambda}{}^{\nu}
-\delta^{\nu}_{\alpha}\Gamma^{\lambda}{}_{\lambda}{}^{\mu})
+k_{2}(\delta^{\nu}_{\alpha}\Gamma^{\mu\lambda}{}_{\lambda}
-\delta^{\mu}_{\alpha}\Gamma^{\nu\lambda}{}_{\lambda})
+k_{3}(\Gamma^{\nu\mu}{}_{\alpha}-\Gamma^{\mu\nu}{}_{\alpha})
\right\},
\end{equation}
where $k_{1}$, $k_{2}$ and $k_{3}$ are the extra added constants.
Inside matter at the origin in Riemann normal coordinates
\begin{eqnarray}
t_{\alpha}{}^{\mu}
&=&\partial_{\nu}U_{\alpha}{}^{[\mu\nu]} \nonumber\\
&=&\frac{1}{3}\sqrt{-g}\left\{
(k_{1}+2k_{2}+3k_{3})R_{\alpha}{}^{\mu}
-(k_{1}+2k_{2})\delta^{\mu}_{\alpha}R\right\} \nonumber\\
&=&2\sqrt{-g}G_{\alpha}{}^{\mu},
\end{eqnarray}
provided that
\begin{eqnarray}
k_{1}+2k_{2}+3k_{3}&=&6, \label{14aNoc2006}\\
k_{1}+2k_{2}&=&3.        \label{14bNoc2006}
\end{eqnarray}
For the ADM mass at the spatial infinity, let's use the
Schwarzschild metric in Cartesian coordinate
\begin{equation}
ds^{2}=-\left(1-\frac{2GM}{r}\right)dt^{2}
+\left(1+\frac{2GM}{r}\right)(dx^{2}+dy^{2}+dz^{2}).
\end{equation}
For the energy-momentum four vector
\begin{equation}
2\kappa{}P_{\alpha}=-\frac{1}{2}U_{\alpha}{}^{[\mu\nu]}\epsilon_{\mu\nu},
\end{equation}
the associated ADM mass energy term is
\begin{equation}
M=-\frac{1}{4\kappa}U_{0}{}^{[\mu\nu]}\epsilon_{\mu\nu}=\frac{1}{2}\left(k_{1}+k_{3}\right)M,
\end{equation}
which generated one more constraint
\begin{equation}
k_{1}+k_{3}=2.  \label{14cNoc2006}
\end{equation}
Consider (\ref{14aNoc2006}), (\ref{14bNoc2006}) and
(\ref{14cNoc2006}), they provide an unique solution
\begin{equation}
k_{1}=k_{2}=k_{3}=1.
\end{equation}
Therefore only the Einstein pseudotensor or other similar ones,
such as the Landau-Lifshitz pseudotensor, have this property.  But
when one examines the gravitational energy for Landau-Lifshitz or
Einstein, they both do not have the good results in vacuum, namely
only the Bel-Robinson tensor \cite{Deser,MTW}.  It seems hopeless
to have the desired result unless one introduces the flat metric
tensor $\eta_{\alpha\beta}$ along with $g_{\alpha\beta}$. The
details will be discussed in the next section.

\section{The calculation results}

\subsection{The Einstein and Landau-Lifshitz pseudotensors}
The superpotentials for the Einstein $_{E}U_{\alpha}{}^{[\mu\nu]}$
and Landau-Lifshitz $_{L}U^{\alpha[\mu\nu]}$, and Bergman-Thomson
$_{B}U^{\alpha[\mu\nu]}$ classical pseudotensors are
\begin{eqnarray}
_{E}U_{\alpha}{}^{[\mu\nu]}&=&-\sqrt{-g}g^{\beta\sigma}\Gamma^{\tau}{}_{\lambda\beta}
\delta_{\tau\sigma\alpha}^{\lambda\mu\nu},\\
_{L}U^{\alpha[\mu\nu]}&=&\sqrt{-g}_{B}U^{\alpha[\mu\nu]}
=gg^{\alpha\beta} g^{\pi\sigma}\Gamma^{\tau}{}_{\lambda\pi}
\delta_{\tau\sigma\beta}^{\lambda\mu\nu}.
\end{eqnarray}
The pseudotensors can be obtained respectively as follows
\begin{equation}
t_{\alpha}{}^{\mu}=\partial_{\nu}U_{\alpha}{}^{[\mu\nu]},\quad
t^{\alpha\mu}=\partial_{\nu}U^{\alpha[\mu\nu]}.
\end{equation}
The result inside matter at the origin for the Einstein and
Bergmann-Thomson pseudotensors are
\begin{equation}
2\kappa{}t_{\alpha}{}^{\beta}(0)
=2G_{\alpha}{}^{\beta}(0)=2\kappa{}T_{\alpha}{}^{\beta}(0),
\end{equation}
so their corresponding energy densities are
\begin{equation}
{\cal{}E}=-t_{0}{}^{0}=-\frac{G_{0}{}^{0}}{\kappa}=-T_{0}{}^{0}=\rho,
\end{equation}
where $\rho$ is the mass-energy density.  This feature is
important because we have to fulfill this basic requirement
according to the equivalent principle. This means the pseudotensor
has to match the energy density inside matter at the origin as the
metric tensor becomes flat.  If the zeroth order is faulty it is
not appropriate  to study the second derivatives which give the
gravitational energy in vacuum. According to \cite{Deser}, the
non-vanishing terms of the Einstein and Landau-Lifshitz
pseudotensors in vacuum are
\begin{eqnarray}
E_{\alpha}{}^{\beta}&=&-2\Gamma_{\lambda\sigma\alpha}\Gamma^{\beta\lambda\sigma}
+\delta^{\beta}_{\alpha}\Gamma_{\lambda\sigma\tau}\Gamma^{\tau\lambda\sigma},\\
L^{\alpha\beta}&=&
\Gamma^{\alpha}{}_{\lambda\sigma}(\Gamma^{\beta\lambda\sigma}-\Gamma^{\lambda\sigma\beta})
-\Gamma_{\lambda\sigma}{}^{\alpha}(\Gamma^{\beta\lambda\sigma}+\Gamma^{\sigma\lambda\beta})
+g^{\alpha\beta}\Gamma_{\lambda\sigma\tau}\Gamma^{\sigma\lambda\tau}.
\end{eqnarray}
The second derivatives of the Einstein pseudotensor \cite{MTW} in
vacuum is
\begin{equation}
\partial^{2}_{\mu\nu}E_{\alpha\beta}
=\frac{1}{9}(4B_{\alpha\beta\mu\nu}-S_{\alpha\beta\mu\nu}).\label{17aNov2006}
\end{equation}
According to \cite{Deser}, a similar calculation gives the
non-vanishing second order contribution for the Landau-Lifshitz
$L_{\alpha\beta}$ classical pseudotensor in vacuum as
\begin{equation}
\partial^{2}_{\mu\nu}L_{\alpha\beta}=
\frac{1}{9}\left(7B_{\alpha\beta\mu\nu}+\frac{1}{2}S_{\alpha\beta\mu\nu}\right).
\label{17bNov2006}
\end{equation}
In order to obtain a positive multiple of $B_{\alpha\beta\mu\nu}$,
they have made the combination
\begin{equation}\label{7thJan2006}
\partial^{2}_{\mu\nu}\left(\frac{1}{2}E_{\alpha\beta}+L_{\alpha\beta}
\right)=B_{\alpha\beta\mu\nu}. \label{17cNov2006}
\end{equation}
This result is good because it guarantees positivity energy for
all reference frames.\\

Looking at (\ref{17aNov2006}), (\ref{17bNov2006}) and
(\ref{17cNov2006}), what do they mean physically?  One way of
getting some physical insight is to integrate over a small
coordinate sphere at time $t=0$, the four momentum is
\begin{eqnarray}
P_{\mu}=(-E,P_{i})=\int t^{0}{}_{\mu}d^{3}x,
\end{eqnarray}
where the energy $E$ should be non-negative.  However $P_{\mu}$ is
not necessarily a proper four vector, because it refers to the
energy-momentum in a unit sphere of $r={\rm{}constant}$ at
$t={\rm{}constant}$ and needs not refer to the energy-momentum in
a unit sphere of $r'={\rm{}constant}$ at $t'={\rm{}constant}$.
These two are not related by a simple Lorentz transformation.
Consider the four momentum to second order for the Einstein
pseudotensor in vacuum within a small sphere, as it appears to the
laboratory frame observer; it is
\begin{eqnarray}
_{E}P_{\mu}&=&\frac{1}{2\kappa}\int\frac{1}{18}(4B^{0}{}_{\mu{}
ij}-S^{0}{}_{\mu ij})x^{i}x^{j}d^{3}x \nonumber \\
&=&-\frac{r^{5}}{540G}\left(2B_{\mu0l}{}^{l}+5T_{\mu0l}{}^{l}
\right)\nonumber\\
&=&-\frac{r^{5}}{540G}
(7E_{ab}E^{ab}-3H_{ab}H^{ab},4\epsilon_{c}{}^{ab}E_{ad}H^{d}{}_{b}),
\end{eqnarray}
since
\begin{equation}
\int{}x^{i}x^{j}d^{3}x=\frac{1}{3}\delta^{ij}\int{}r^{2}d^{3}x=\frac{4\pi}{15}r^{5}\delta^{ij}.
\end{equation}
Similarly for the Landau-Lifshitz pseudotensor
\begin{eqnarray}
_{L}P_{\mu}&=&\frac{1}{2\kappa}\int\frac{1}{18}\left(7B^{0}{}_{\mu{}ij}
+\frac{1}{2}S^{0}{}_{\mu ij}\right)x^{i}x^{j}d^{3}x \nonumber \\
&=&-\frac{r^{5}}{1080G}\left(7B_{\mu0l}{}^{l}-5T_{\mu0l}{}^{l}\right) \nonumber\\
&=&-\frac{r^{5}}{540G}(E_{ab}E^{ab}+6H_{ab}H^{ab},7\epsilon_{c}{}^{ab}E_{ad}H^{d}{}_{b}).
\end{eqnarray}
Combining these two pseudotensors, the properly normalized
superpotential becomes
\begin{equation}
_{EL}U^{\alpha[\mu\nu]}
=\frac{2}{3}\left\{\frac{1}{2}\eta^{\alpha\lambda}{}_{E}U_{\lambda}{}^{[\mu\nu]}
+{}_{B}U^{\alpha[\mu\nu]}\right\}.
\end{equation}
The gravitational energy-momentum within a small sphere is
\begin{eqnarray}\label{12aJune2006}
_{EL}P_{\mu}
&=&\frac{2}{3}\left(\frac{1}{2}{}_{E}P_{\mu}+{}_{B}P_{\mu}\right) \nonumber\\
&=&-\frac{r^{5}}{180G}B_{\mu{}0l}{}^{l} \nonumber\\
&=&-\frac{r^{5}}{180G}(E_{ab}E^{ab}+H_{ab}H^{ab},2\epsilon_{c}{}^{ab}E_{ad}H^{d}{}_{b}).
\end{eqnarray}
This result gives the coefficient of the energy-momentum density
should be.  In other words, the coefficient of the Bel-Robinson
tensor in vacuum in a holonomic frame is
\begin{equation}
\partial^{2}_{\mu\nu}t_{\alpha}{}^{\beta}
=\frac{2}{3}B_{\alpha}{}^{\beta}{}_{\mu\nu}.
\end{equation}

\subsection{The modification of the Einstein pseudotensor}

The Einstein pseudotensor offers a good foundation for the
gravitational energy expression.  One may wonder that if this
pseudotensor can be modified so that it has a nice positive
gravitational energy result.  Based on the simple and natural
boundary conditions, analogy with the Dirichlet or Neumann
boundary conditions, one of the Chen-Nester four quasilocal
expressions can satisfy this requirement \cite{So2}.\\

According to \cite{So2}, there is a two parameters set for the
modified Chen-Nester quasilocal boundary expressions.  There are
infinite number of solutions to achieve a positive Bel-Robinson
tensor. The quasilocal boundary expressions, in compact form, are
\begin{eqnarray}\label{9June2006}
2\kappa{\cal{}B}_{c_{1},c_{2}}(N)
&=&2\kappa{\cal{}B}_{p}(N)+c_{1}i_{N}\Delta{}\Gamma^{\alpha}{}_{\beta}
\wedge\Delta\eta_{\alpha}{}^{\beta}
-c_{2}\Delta\Gamma^{\alpha}{}_{\beta}
\wedge{}i_{N}\Delta\eta_{\alpha}{}^{\beta} \nonumber\\
&=&-\frac{1}{2}N^{\alpha}\left\{ _{E}U_{\alpha}{}^{[\mu\nu]}
+c_{1}\sqrt{-g}h^{\lambda\pi}\Gamma^{\sigma}{}_{\alpha\pi}
\delta^{\mu\nu}_{\lambda\sigma}
+c_{2}\sqrt{-g}h^{\beta\sigma}\Gamma^{\tau}{}_{\lambda\beta}
\delta^{\lambda\mu\nu}_{\tau\sigma\alpha}
\right\}\epsilon_{\mu\nu},
\end{eqnarray}
where $c_{1},c_{2}\in\Re$ and
$h_{\alpha\beta}:=g_{\alpha\beta}-\eta_{\alpha\beta}$. When
$(c_{1},c_{2})=(0,0),~(0,1),~(1,0)$ and $(1,1)$, this recovers the
original Chen-Nester four expressions.  From (\ref{9June2006}),
the superpotential can be extracted as
\begin{equation}
U_{\alpha}{}^{[\mu\nu]} ={}_{E}U_{\alpha}{}^{[\mu\nu]}
+c_{1}\sqrt{-g}h^{\lambda\pi}\Gamma^{\sigma}{}_{\alpha\pi}
\delta^{\mu\nu}_{\lambda\sigma}
+c_{2}\sqrt{-g}h^{\beta\sigma}\Gamma^{\tau}{}_{\lambda\beta}
\delta^{\lambda\mu\nu}_{\tau\sigma\alpha}.
\end{equation}
This superpotential looks like a modification of the Freud
superpotential.  Note that the $h\Gamma$ terms do not affect the
results inside matter and at spatial infinity, but they do affect
second order vacuum value.  As mentioned before, the pseudotensor
can be obtained as
\begin{equation}
t_{\alpha}{}^{\mu}=\partial_{\nu}U_{\alpha}{}^{[\mu\nu]}.
\end{equation}
For all $c_{1}$ and $c_{2}$, inside matter at the origin the
result is as desired:
\begin{equation}
2\kappa{}t_{\alpha}{}^{\beta}(0)
=2G_{\alpha}{}^{\beta}(0)=2\kappa{}T_{\alpha}{}^{\beta}(0),
\end{equation}
and the corresponding energy density is
\begin{equation}
{\cal{}E}=-t_{0}{}^{0}=-\frac{G_{0}{}^{0}}{\kappa}=-T_{0}{}^{0}=\rho.
\end{equation}
The second derivatives in vacuum are
\begin{equation}
\partial^{2}_{\mu\nu}t_{\alpha}{}^{\beta}(0)
=\frac{1}{9}\left\{
(4+c_{1}-5c_{2})B_{\alpha}{}^{\beta}{}_{\mu\nu}
-(1-2c_{1}+c_{2})S_{\alpha}{}^{\beta}{}_{\mu\nu}
+(c_{1}-3c_{2})(Y_{\alpha}{}^{\beta}{}_{\mu\nu}+9T_{\alpha}{}^{\beta}{}_{\mu\nu})
\right\},
\end{equation}
and the gravitational energy-momentum is
\begin{eqnarray}\label{12June2006}
P_{\mu}&=&\frac{1}{2\kappa}\int\frac{1}{18}\left\{
\begin{array}{cccc}
 (4+c_{1}-5c_{2})B^{0}{}_{\mu{}ij}
-(1-2c_{1}+c_{2})S^{0}{}_{\mu{}ij}\\
+(c_{1}-3c_{2})(Y^{0}{}_{\mu{}ij}+9T^{0}{}_{\mu{}ij})
\quad\quad\quad\quad\quad\quad\quad\\
\end{array} \right\}
x^{i}x^{j}d^{3}x \nonumber \\
&=&-\frac{r^{5}}{540G}\left\{
(2+c_{1}-4c_{2})B_{\mu0l}{}^{l}+5(1-c_{1}-2c_{2})T_{\mu0l}{}^{l}
\right\}.
\end{eqnarray}
In order to obtain the positive gravitational energy, the
coefficients of $B_{\mu{}0l}{}^{l}$ should be positive and that of
$T_{\mu{}0l}{}^{l}$ should vanish.  Namely
\begin{equation}
2+c_{1}-4c_{2}>0,\quad\quad 1-c_{1}-2c_{2}=0,
\end{equation}
or equivalently
\begin{equation}
c_{1}>0,  \quad\quad{}  (c_{1},c_{2})=\left(c_{1},
\frac{1-c_{1}}{2} \right).
\end{equation}
Then rewriting (\ref{12June2006})
\begin{eqnarray}
P_{\mu}&=&-\frac{c_{1}r^{5}}{180G}B_{\mu0l}{}^{l}\nonumber\\
&=&-\frac{c_{1}r^{5}}{180G}
(E_{ab}E^{ab}+H_{ab}H^{ab},2\epsilon_{c}{}^{ab}E_{ad}H^{d}{}_{b}).
\end{eqnarray}
There are an infinite number of solutions since there is a
parameter $c_{1}$ which one can tune.  Different solutions
represent different boundary conditions which is explained in
\cite{So2}. However there is one solution with positive energy and
a simple boundary condition which is when $(c_{1},c_{2})=(1,0)$ in
(\ref{9June2006}).  It reduces to one of the original Chen-Nester
expressions that they called ${\cal{}B}_{c}(N)$.  In detail
\begin{equation}
2\kappa{\cal{}B}_{c}(N)=i_{N}\Gamma^{\alpha}{}_{\beta}\wedge\Delta\eta_{\alpha}{}^{\beta}
+\Delta\Gamma^{\alpha}{}_{\beta}\wedge{}i_{N}\eta_{\alpha}{}^{\beta},
\end{equation}
the corresponding superpotential is
\begin{equation}\label{9aJune2006}
{\cal{}U}_{\alpha}{}^{[\mu\nu]}={}_{E}U_{\alpha}{}^{[\mu\nu]}
+\sqrt{-g}h^{\lambda\pi}\Gamma^{\sigma}{}_{\alpha\pi}
\delta^{\mu\nu}_{\lambda\sigma}.
\end{equation}
This modification is not just achieving the Bel-Robinson tensor by
using an artificial combination, it comes from a simple Lagrangian
derivation and a Hamiltonian analysis beginning from the least
action principle. Inside matter at the origin and the second
derivatives of the small sphere in vacuum results respectively are
\begin{eqnarray}
2\kappa{}t_{\alpha}{}^{\beta}(0)
&=&2G_{\alpha}{}^{\beta}(0)~=~2\kappa{}T_{\alpha}{}^{\beta}(0),\\
\partial^{2}_{\mu\nu}t_{\alpha\beta}(0)
&=&\frac{1}{9}(5B_{\alpha\beta\mu\nu}+S_{\alpha\beta\mu\nu}
+Y_{\alpha\beta\mu\nu}+9T_{\alpha\beta\mu\nu}).
\end{eqnarray}
The result inside matter is good.  But the second derivatives of
the pseudotensor does not give only the Bel-Robinson tensor.
However after integration, the gravitational energy-momentum turns
out to be what we expect:
\begin{eqnarray}
P_{\mu}&=&\frac{1}{2\kappa}\int\frac{1}{18}\left(5B^{0}{}_{\mu{}ij}
+S^{0}{}_{\mu{}ij}+Y^{0}{}_{\mu{}ij}+9T^{0}{}_{\mu{}ij}\right)x^{i}x^{j}
d^{3}x \nonumber \\
&=&-\frac{r^{5}}{180G}B_{\mu0l}{}^{l} \nonumber\\
&=&-\frac{r^{5}}{180G}(E_{ab}E^{ab}+H_{ab}H^{ab},2\epsilon_{c}{}^{ab}E_{ad}H^{d}{}_{b}),
\end{eqnarray}
which is the same as (\ref{12aJune2006}).  This result shows that
the vector $P_{\mu}$ is future pointing and non-spacelike.

\subsection{The modification of the Landau-Lifshitz pseudotensor}
The superpotentials for the Papapetrou classical pseudotensor is
\begin{equation}
_{P}H^{[\mu\nu][\alpha\beta]}
=-\sqrt{-g}g^{ma}\eta^{nb}\delta_{ab}^{\nu\mu}\delta_{mn}^{\alpha\beta},
\end{equation}
or equivalently
\begin{eqnarray}
_{P}U^{\alpha[\mu\nu]}
&=&\partial_{\beta}\left(_{P}H^{[\mu\nu][\alpha\beta]}\right) \nonumber\\
&=&_{B}U^{\alpha[\mu\nu]}-\sqrt{-g}\left(
g^{\lambda\sigma}h^{\pi\beta}\Gamma^{\alpha}{}_{\lambda\pi}\delta^{\nu\mu}_{\sigma\beta}
+g^{\alpha\beta}h^{\pi\sigma}\Gamma^{\tau}{}_{\lambda\pi}\delta^{\lambda\mu\nu}_{\tau\sigma\beta}
\right),
\end{eqnarray}
At this point, this superpotential looks like a modification of
the Bergmann-Thomson or Landau-Lifshitz pseudotensor.  Once again,
the $h\Gamma$ terms do not affect the results inside matter and at
spatial infinity, they would contribute however to the second
order vacuum value. As mentioned before, the pseudotensor can be
treated as
\begin{equation}
t^{\alpha\mu}=\partial_{\nu}U^{\alpha[\mu\nu]}.
\end{equation}
Inside matter at the origin and the second derivatives in vacuum
\cite{So1} are respectively
\begin{eqnarray}
2\kappa{}t_{\alpha}{}^{\beta}(0)
&=&2G_{\alpha}{}^{\beta}(0)~=~2\kappa{}T_{\alpha}{}^{\beta}(0),\\
\partial^{2}_{\mu\nu}P_{\alpha\beta}(0)&=&\frac{2}{9}(4B_{\alpha\mu\nu}-S_{\alpha\mu\nu}
-Y_{\alpha\mu\nu}-9T_{\alpha\beta\mu\nu}).
\end{eqnarray}
The result inside matter is good but the second derivative term
does not give the desired result. However after integration, the
gravitational energy-momentum becomes good:
\begin{eqnarray}
P_{\mu} &=&\frac{1}{2\kappa}\int\frac{1}{9}\left(
4B^{0}{}_{\mu{}ij}-S^{0}{}_{\mu{}ij}
-Y^{0}{}_{\mu{}ij}-9T^{0}{}_{\mu{}ij}\right)x^{i}x^{j}d^{3}x \nonumber\\
&=&-\frac{r^{5}}{180G}B_{\mu{}0l}{}^{l} \nonumber\\
&=&-\frac{r^{5}}{180G}(E_{ab}E^{ab}+H_{ab}H^{ab},2\epsilon_{c}{}^{ab}E_{ad}H^{d}{}_{b}).
\end{eqnarray}
Once again, it is the same as (\ref{12aJune2006}) and, as said
previously, the vector $P_{\mu}$ is future pointing and
non-spacelike.

\section{Conclusion}
The Einstein and Landau-Lifshitz pseudotensors do not give a
positive gravitational energy in vacuum in terms to second order.
In order to obtain the expected result of the Bel-Robinson tensor
in vacuum, one may consider the modification of these two
pseudotensors.  One of the possible modifications of the Einstein
pseudotensor turns out to be one of the Chen-Nester quasilocal
expressions.  For the modification of the Landau-Lifshitz
pseudotensor, one of the choices is the Papapretou pseudotensor.
There are two points that need to be emphasized, they both have
the same magnitude of the $B_{\mu{}0l}{}^{l}$ term and they both
correspond to natural boundary conditions.


\end{document}